\documentclass[aps,prl,twocolumn,showpacs,superscriptaddress,tightenlines,floatfix]{revtex4}

\usepackage{amsmath}
\usepackage{amssymb}
\usepackage{graphicx}
\usepackage{epsfig}
\usepackage{color}
\usepackage{multirow}
\usepackage{rjwmath}
\usepackage[mathcal]{eucal}

\begin{document}



\title{ \quad\\[1.0cm] Study of the Hadronic Transitions $\Upsilon$(2S)$\rightarrow (\eta,\pi^0)\Upsilon$(1S) at Belle}

\affiliation{University of the Basque Country UPV/EHU, Bilbao}
\affiliation{Budker Institute of Nuclear Physics SB RAS and Novosibirsk State University, Novosibirsk 630090}
\affiliation{Faculty of Mathematics and Physics, Charles University, Prague}
\affiliation{University of Cincinnati, Cincinnati, Ohio 45221}
\affiliation{II. Physikalisches Institut, Georg-August-Universit\"at G\"ottingen, G\"ottingen}
\affiliation{Hanyang University, Seoul}
\affiliation{University of Hawaii, Honolulu, Hawaii 96822}
\affiliation{High Energy Accelerator Research Organization (KEK), Tsukuba}
\affiliation{IKERBASQUE, Bilbao}
\affiliation{Indian Institute of Technology Guwahati, Guwahati}
\affiliation{Institute of High Energy Physics, Chinese Academy of Sciences, Beijing}
\affiliation{Institute of High Energy Physics, Vienna}
\affiliation{Institute of High Energy Physics, Protvino}
\affiliation{INFN - Sezione di Torino, Torino}
\affiliation{Institute for Theoretical and Experimental Physics, Moscow}
\affiliation{J. Stefan Institute, Ljubljana}
\affiliation{Kanagawa University, Yokohama}
\affiliation{Institut f\"ur Experimentelle Kernphysik, Karlsruher Institut f\"ur Technologie, Karlsruhe}
\affiliation{Korea Institute of Science and Technology Information, Daejeon}
\affiliation{Korea University, Seoul}
\affiliation{Kyungpook National University, Taegu}
\affiliation{\'Ecole Polytechnique F\'ed\'erale de Lausanne (EPFL), Lausanne}
\affiliation{Faculty of Mathematics and Physics, University of Ljubljana, Ljubljana}
\affiliation{Luther College, Decorah, Iowa 52101}
\affiliation{University of Maribor, Maribor}
\affiliation{Max-Planck-Institut f\"ur Physik, M\"unchen}
\affiliation{Graduate School of Science, Nagoya University, Nagoya}
\affiliation{Kobayashi-Maskawa Institute, Nagoya University, Nagoya}
\affiliation{National Central University, Chung-li}
\affiliation{National United University, Miao Li}
\affiliation{Department of Physics, National Taiwan University, Taipei}
\affiliation{H. Niewodniczanski Institute of Nuclear Physics, Krakow}
\affiliation{Nippon Dental University, Niigata}
\affiliation{Niigata University, Niigata}
\affiliation{Osaka City University, Osaka}
\affiliation{Pacific Northwest National Laboratory, Richland, Washington 99352}
\affiliation{Research Center for Electron Photon Science, Tohoku University, Sendai}
\affiliation{University of Science and Technology of China, Hefei}
\affiliation{Seoul National University, Seoul}
\affiliation{Sungkyunkwan University, Suwon}
\affiliation{School of Physics, University of Sydney, NSW 2006}
\affiliation{Tata Institute of Fundamental Research, Mumbai}
\affiliation{Excellence Cluster Universe, Technische Universit\"at M\"unchen, Garching}
\affiliation{Toho University, Funabashi}
\affiliation{Tohoku Gakuin University, Tagajo}
\affiliation{Tohoku University, Sendai}
\affiliation{Department of Physics, University of Tokyo, Tokyo}
\affiliation{Tokyo Institute of Technology, Tokyo}
\affiliation{Tokyo Metropolitan University, Tokyo}
\affiliation{Tokyo University of Agriculture and Technology, Tokyo}
\affiliation{CNP, Virginia Polytechnic Institute and State University, Blacksburg, Virginia 24061}
\affiliation{Wayne State University, Detroit, Michigan 48202}
\affiliation{Yamagata University, Yamagata}
\affiliation{Yonsei University, Seoul}
  \author{U.~Tamponi}\affiliation{INFN - Sezione di Torino, Torino} 
  \author{R.~Mussa}\affiliation{INFN - Sezione di Torino, Torino} 
  \author{I.~Adachi}\affiliation{High Energy Accelerator Research Organization (KEK), Tsukuba} 
  \author{H.~Aihara}\affiliation{Department of Physics, University of Tokyo, Tokyo} 
  \author{D.~M.~Asner}\affiliation{Pacific Northwest National Laboratory, Richland, Washington 99352} 
  \author{V.~Aulchenko}\affiliation{Budker Institute of Nuclear Physics SB RAS and Novosibirsk State University, Novosibirsk 630090} 
  \author{T.~Aushev}\affiliation{Institute for Theoretical and Experimental Physics, Moscow} 
  \author{A.~M.~Bakich}\affiliation{School of Physics, University of Sydney, NSW 2006} 
  \author{M.~Barrett}\affiliation{University of Hawaii, Honolulu, Hawaii 96822} 
  \author{B.~Bhuyan}\affiliation{Indian Institute of Technology Guwahati, Guwahati} 
% ...
\collaboration{The Belle Collaboration}


\begin{abstract}
We study the rare hadronic transitions 
$\Upsilon(2S)\rightarrow  \Upsilon(1S)\eta$ and
$\Upsilon(2S)\rightarrow  \Upsilon(1S)\pi^0$ using a sample of 158 $\times
10^6$  $\Upsilon(2S)$ decays collected with the Belle
detector at the KEKB asymmetric-energy $e^+ e^-$ collider.
We measure the ratios of branching fractions (${\mathcal B}$)
$\frac{{\mathcal B}(\Upsilon(2S)\rightarrow\Upsilon(1S)\eta)}{{\mathcal B}(\Upsilon(2S)\rightarrow\Upsilon(1S)\pi^+\pi^-)}$ =
(1.99$\pm$0.14 (stat) $\pm$0.11 (syst)) $\times 10^{-3}$ and 
 $\frac{{\mathcal B}(\Upsilon(2S)\rightarrow\Upsilon(1S)\pi^0)}{{\mathcal B}(\Upsilon(2S)\rightarrow\Upsilon(1S)\pi^+\pi^-)} < 2.3 \times 10^{-4}$ at the $90\%$ confidence level (CL).
Assuming the value ${\mathcal B}(\Upsilon(2S) \rightarrow  \Upsilon(1S)\pi^-\pi^+)$ = (17.92$\pm$0.26)\%, we obtain 
$
{\mathcal B}(\Upsilon(2S)\rightarrow\Upsilon(1S)\eta) 
= (3.57 \pm 0.25 ({\rm stat})\  \pm 0.21 ({\rm syst}))\times 10^{-4} 
$
and 
$
{\mathcal B}(\Upsilon(2S)\rightarrow\Upsilon(1S)\pi^0) < 4.1\times 10^{-5}\  ({\rm 90\%\ CL}).
$

\end{abstract}

\maketitle

\tighten

{\renewcommand{\thefootnote}{\fnsymbol{footnote}}}
\setcounter{footnote}{0}

In recent years, hadronic transitions between quarkonia have led to 
an impressive series of discoveries \cite{Brambilla:2010cs}: $X$(3872), $Y$(4260), as well
as $h_c$ and $h_b$ were observed in transitions either from or to the
$\psi$ and $\Upsilon$ states. The phenomenology of these transitions is commonly
described with the QCD Multipole Expansion formalism (QCDME) \cite{Gottfried:1977gp,Yan:1980uh},
which allows one to classify the transitions in a series of chromoelectric
and chromomagnetic multiplets.
In particular, theoretical predictions for  $\eta$ and $\pi^0$ transitions   
\cite{Voloshin:2007dx,Kuang:2006me} among
states are being challenged by experimental measurements. 
The  $\eta$ and $\pi^0$ transitions between vector bottomonia should be mediated either by 
two M1 gluons or by one E1 and one M2 gluon: both cases imply a spin flip of the $b$ quark.
The corresponding amplitude should scale as $1/m_b$, and its
measurement yields information about the chromomagnetic moment of the
$b$ quark.

By scaling from the $\psi(2S)\to J/\psi \eta$ transition, one expects a transition width of: 
$\Gamma[\Upsilon(2S) \to  \Upsilon(1S)\eta ] = 0.0025 \times \Gamma[\psi(2S)\to J/\psi \eta]$,
 and therefore a ratio of branching fractions  ${\mathcal R}_{\eta,\pi^+\pi^-} =\frac{{\mathcal B}(\Upsilon(2S)\rightarrow\Upsilon(1S)\eta)}{{\mathcal B}(\Upsilon(2S)\rightarrow\Upsilon(1S)\pi^+\pi^-)}  \approx 4.4\times 10^{-3}$ \cite{He:2008xk}.
From \cite{Voloshin:2007dx}, one can calculate ${\mathcal R}_{\eta,\pi^+\pi^-} = 2.3 \times 10^{-3}$ assuming the $b$ quark mass to be $m_b = 4.67$ GeV/c$^2$  \cite{Nakamura:2010zzi}; the prediction from \cite{Kuang:2006me} is ${\mathcal R}_{\eta,\pi^+\pi^-} = 1.7 \times 10^{-3}$.
A further suppression is expected for the $\pi^0$ transition, which violates isospin; 
here, theory predicts  $\Gamma[\Upsilon(2S) \to  \Upsilon(1S)\pi^0 ]  =
 0.16 \times \Gamma[\Upsilon(2S) \to  \Upsilon(1S)\eta ]$ . 
The $\Upsilon(2,3,4S)\rightarrow \Upsilon(1S)\eta$ transitions 
have been studied by BaBar \cite{Babar:2008bv, Babar:2011wj} and 
CLEO \cite{He:2008xk}; the measured branching fractions are either unexpectedly
 large ($\Upsilon(4S)$) or too small ($\Upsilon(2S)$ and $\Upsilon(3S)$). 
The parameters of the quark wave functions must be changed by
more than 15\% in order to account for these discrepancies \cite{Simonov:2008sw}.
Searches for the $\pi^0$ transitions have only yielded upper
limits \cite{He:2008xk,Babar:2011wj}.

We report here a new measurement of the transition
$\Upsilon(2S) \rightarrow  \Upsilon(1S)\eta$ and a search for
$\Upsilon(2S) \rightarrow  \Upsilon(1S)\pi^0$ using the Belle detector at the KEKB e$^+$e$^-$ collider \cite{kekb}. 
The $\Upsilon(1S)$ is reconstructed in both the $e^+e^-$ and $\mu^+\mu^-$ 
 decay modes; we reconstruct the $\eta$ meson via its decay to $\gamma\gamma$ 
and $\pi^-\pi^+\pi^0$, and the $\pi^0$ in the $ \gamma\gamma$ final state.
As a normalization sample, we reconstruct the dominant transition 
$\Upsilon(2S) \rightarrow  \Upsilon(1S)\pi^-\pi^+$, which has a branching fraction of 
(17.92$\pm$0.26)\% \cite{Nakamura:2010zzi}.
The data sample for this
analysis includes an integrated luminosity of 24.7 fb$^{-1}$ at the $\Upsilon(2S)$ resonance peak, corresponding to 
(158$\pm$4)$\times 10^6$ $\Upsilon(2S)$ decays, and an additional  1.7 fb$^{-1}$
at $\sqrt{s}$=9.993 GeV used to study
the QED continuum backgrounds.

 The Belle detector is described  in detail elsewhere~\cite{Abashian:2000cg}.
 Here, we summarize the features that are relevant to the current analysis.
The momentum of each charged track is measured using 
a four-layer double-sided Silicon Vertex Detector  and a 50 layer
Central Drift Chamber, embedded in a 1.5 T magnetic field, which allows tracking of charged
particles with transverse momentum as low as 50 MeV/c.
Electrons and photons are detected  in a large array of 
CsI(Tl) crystals (ECL) also located inside the solenoid coil. 
An iron flux return located outside 
the coil is instrumented to detect $K_L^0$ mesons and to identify
muons.  

Monte Carlo (MC) samples of the signal and of the dominant peaking backgrounds were generated
using EvtGen \cite{Lange:2001uf}.
Dipion transitions were generated
assuming that the amplitude is dominated by the $S$-wave contribution. 
QED continuum processes $e^+ e^- \to (e^+e^-$ or $\mu^+\mu^-) + n\gamma$ were generated using KKMC \cite{Jadach:1999vf}.
The detector response was simulated using GEANT3 
\cite{Geant3}, and beam backgrounds were accounted 
for using random triggers taken during each period of data taking. 
Final state radiation effects are accounted for by using PHOTOS \cite{Barberio:1993qi}
 in EvtGen simulations. 

Charged tracks with momentum $p^*$ in the center-of-mass frame of the colliding $e^+e^-$ pair (CM frame) greater than 4 GeV/c are selected as
candidate leptons from $\Upsilon(1S)$ decay. In the following text, all the quantities computed in the CM frame are denoted with an asterisk.
Electrons and muons are identified by the ratio $R_{e, \mu}$ between the electron (muon) hypothesis likelihood and the hadronic hypothesis likelihood.
 A track is identified as a lepton if  $R_{e}$ or $R_{\mu}$  is above a threshold value of  0.2 and then as a muon if $R_{\mu} > R_{e}$. 
 The identification efficiency is $93.2\%$ for electrons and $92.6 \%$ for muons.
Pairs of lepton candidates with opposite charge and an invariant mass
 in the range 9.0 GeV/c$^2 < $ M($\ell^+\ell^-$) $<$ 9.8 GeV/c$^2$ are then selected for further analysis.

In order to reduce the effect of final state radiation (FSR) and bremsstrahlung,
 the momentum of all photons detected in the ECL within 200 mrad 
 of each leptonic track is added to its momentum.  
 A mass-constrained kinematic fit performed on the $\Upsilon(1S)$ candidate lepton pair is required to have
 a confidence level CL$_{1S} > 10^{-5}$.

A requirement on  the polar angle in the CM frame of the $e^-$ track with respect to the beam direction, 
$\cos(\theta^*_{e^-})<0.5$, is imposed in the reconstruction of the $\eta \to \gamma\gamma, \Upsilon(1S) \to e^+e^-$ final state in order to suppress singly- or doubly-radiative Bhabha events, which represent the dominant QED background for this channel. The  Bhabha requirement is not  included 
 in the $\Upsilon(2S) \to  \Upsilon(1S)\pi^0$  analysis since the CL$_{1S}$ requirement provides sufficient
suppression.

Dipion candidates
used both for $\eta \to \pi^+\pi^-\pi^0$ and the $\Upsilon(2S)\to\Upsilon(1S)\pi^+\pi^-$ transition are
composed of oppositely charged pairs of tracks, each of which has a
distance of closest approach of less than 1.5 cm (0.5 cm) in the
axial (transverse) direction relative to the beam line.  The cosine of the angle in the CM frame between these tracks is
required to be less than 0.6 in order to reject events with photons
that convert in the inner detectors.

In the $\eta \to \pi^+\pi^-\pi^0$ analysis, each photon produced in the $ \pi^0$ decay
must have  $E_{\rm lab} >57$ MeV, the optimal threshold for the rejection of photons arising from beam background, and $E^* <220$ MeV. 
The $\pi^0$ candidate is then selected as the $\gamma\gamma$ pair with invariant
mass closest to the nominal $\pi^0$ mass \cite{Nakamura:2010zzi}. The threshold values are optimized by maximizing the figure of merit, defined as $FoM = \frac{s}{\sqrt{s+b}}$, where $s$ is the signal yield and $b$ is the background contribution.

In the $\eta\to\gamma\gamma$ analysis,  photons with  $180$ MeV $< E^{*}< 360 $ MeV 
are subjected to the requirement on the opening angle in the CM frame of $\cos\theta^{*}_{\gamma\gamma}<-0.88$ in order to reject combinatorial background.
Events with more than one $\eta$ candidate that satisfies these conditions are found to be a negligible fraction of the total Monte Carlo sample, and are rejected without introducing any further selection.

The planarity of the event is exploited in order to select the $\Upsilon(2S) \to  \Upsilon(1S)\pi^0$ decay. 
We select the pair of photons with 
 CM momentum  $\vect{p}^*_{\gamma\gamma}$ that minimizes  the scalar product
 $ (\vect{p}^*_{\gamma\gamma} \cdot \hat{u}_{\ell^+\ell^-} )/|{p^*_{\gamma\gamma}}| $ 
 (where  $\hat{u}_{\ell^+\ell^-} $ is a vector normal to the plane formed by the dilepton pair in the CM frame) as the $\pi^0 \to \gamma\gamma$ candidate.

After  $\eta$ or $\pi^0$ selection, the $\Upsilon(1S)$ and $\gamma\gamma$ or $\pi^+\pi^-\pi^0$ are subject to a
second kinematic fit, constraining them to have the $\Upsilon(2S)$ invariant
mass.  The minimum confidence level CL$_{2S}$ is optimized for each decay
mode using MC samples.  For the $\eta \to \gamma\gamma$, CL$_{2S}$ is required
to be greater than $6\times10^{-4}$ $(2\times10^{-3})$ for $\mu^+\mu^- (e^+e^-)$ events;
for $\eta\to\pi^+\pi^-\pi^0$, the event is accepted if the fit converges.  Finally, in the analysis of $\Upsilon(2S)\to \Upsilon(1S)\pi^0$,
CL$_{2S}$ must be greater than $10^{-5}$ for both $\Upsilon(1S)$ decay modes.


Since the signal events are fully reconstructed and  the total momentum of all the charged tracks and photons in the CM frame is expected to be close to zero,  a selection on $p^*_{\rm tot} = |\vect{p_{\Upsilon(1S)}}+\vect{p_{\eta, \pi^0}}|$ is imposed, requiring $p^*_{\rm tot}< $0.07 GeV/c in $\Upsilon(1S)\to e^+e^-, \eta \to \gamma\gamma$ events, which are more contaminated by radiative Bhabha, and  $p^*_{\rm tot}< $0.1 GeV/c when investigating the other modes. 

The requirements described above result in an almost complete rejection of the QED and $\Upsilon(2S) \to \Upsilon(1S) \pi^0\pi^0$ backgrounds. 

The $\Upsilon(2S) \to \chi_{bJ}(1P)\gamma_1  \to  \Upsilon(1S)\gamma_1 \gamma_2$ decay has
the same event topology as $\eta\to\gamma\gamma$ and $\pi^0$ transitions.
The kinematic limit of the less energetic photon, which arises from
 $\Upsilon(2S) \to  \chi_{bJ}(1P) \gamma_1$ decay, is 162 MeV in the CM frame.
Therefore, this background is completely rejected for $\eta\to\gamma\gamma$ 
by the photon energy requirement mentioned earlier.
In the search for the $\pi^0$ transition, this background is still
larger than expected signal; hence, we require the energy of the
less energetic photon to be higher than 170 MeV.
This requirement rejects $99.5\%$ of the $\chi_{bJ}(1P)$ background and retains  $34\%$ of the signal events.


The $\pi^+\pi^-$ transition represents a significant source of background
only for the $\eta \to \pi^+\pi^-\pi^0$ channel; in this case, we require $\Delta M = M(\pi^+ \pi^- \ell^+ \ell^-) - M(\ell^+ \ell^-) < 0.44 $GeV/c$^2$, since this observable peaks at $\Delta M = M(\Upsilon(2S))- M(\Upsilon(1S)) = {\rm 0.56~GeV/c}^2$ in $\pi^+\pi^-$ events; this cut rejects $99.92\%$ of the background and retains $99.1\%$ of the signal. 


Signal
efficiencies for the various final states are summarized in Table \ref{tab:eff_signal}.
The efficiency in the $e^+e^-\gamma\gamma$ mode significantly differs from the one in $\mu^+\mu^-\gamma\gamma$  since this channel is affected by a Bhabha veto included at the trigger level. A trigger simulation is used in order to account for this effect.

\begin{figure}[htbp]
  \begin{center}
   \begin{tabular}{c}
    \resizebox{.99\columnwidth}{!}{\includegraphics{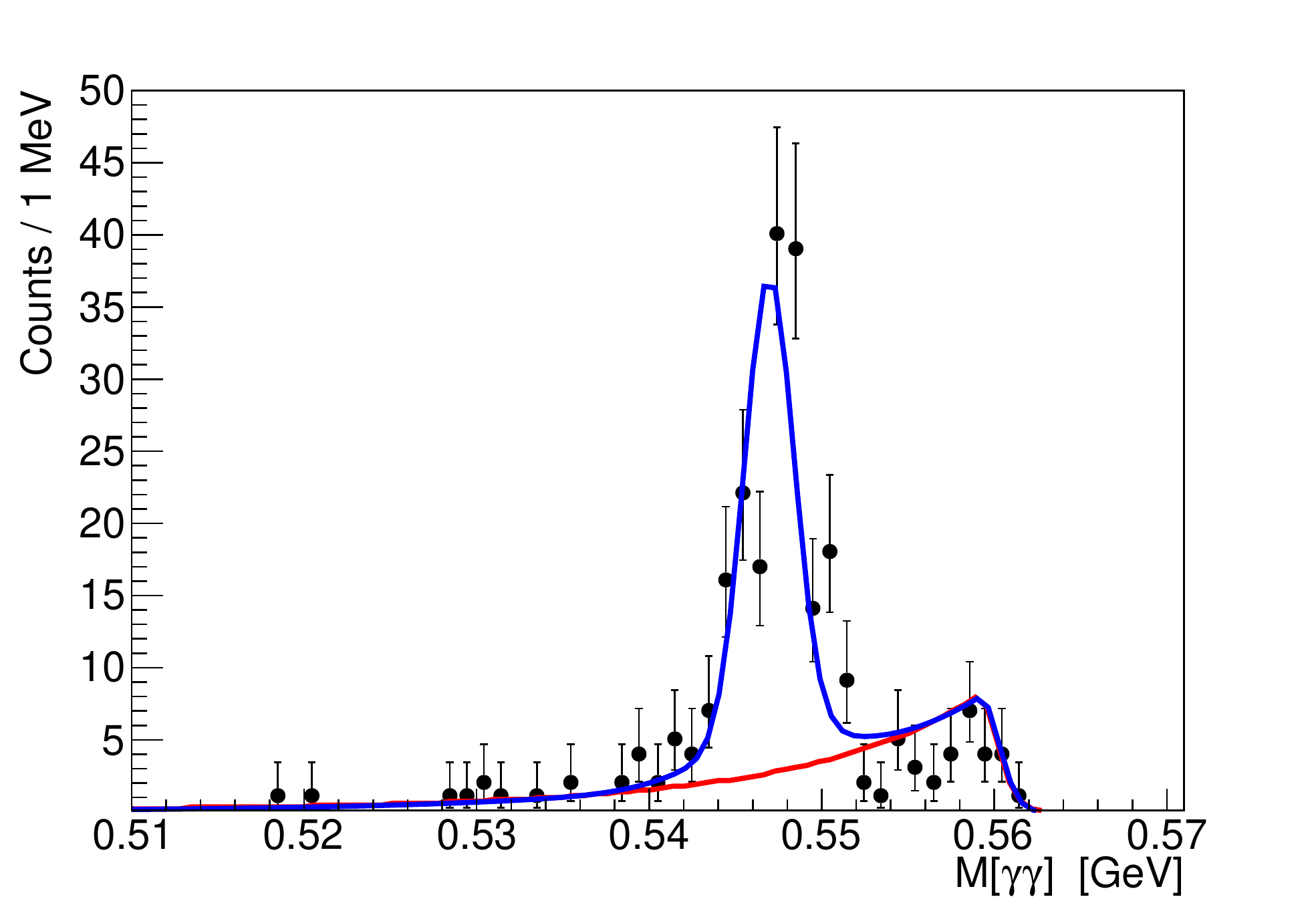}} \\
    \resizebox{.99\columnwidth}{!}{\includegraphics{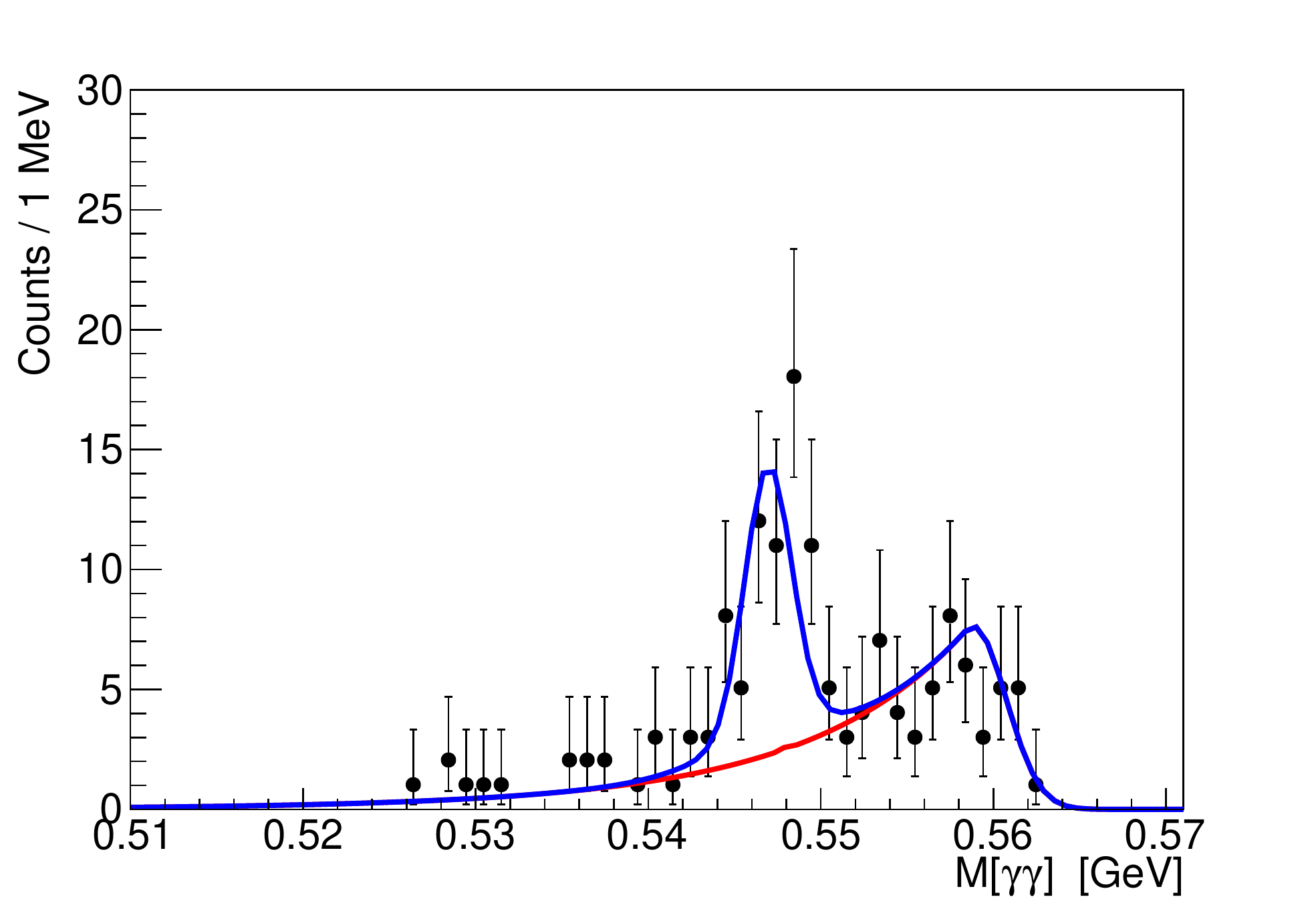}}  \\
    \resizebox{.99\columnwidth}{!}{\includegraphics{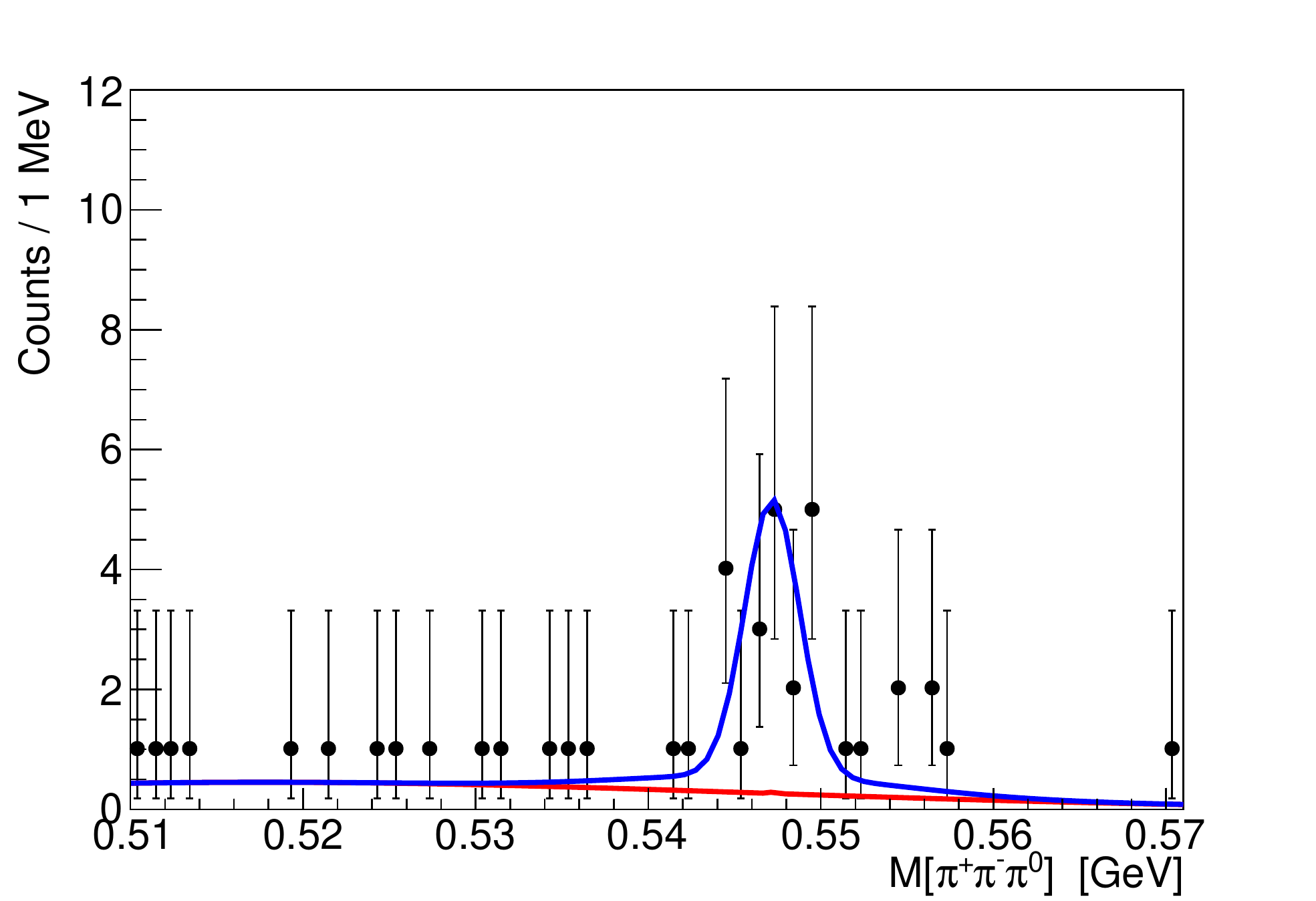}} \\
    \resizebox{.99\columnwidth}{!}{\includegraphics{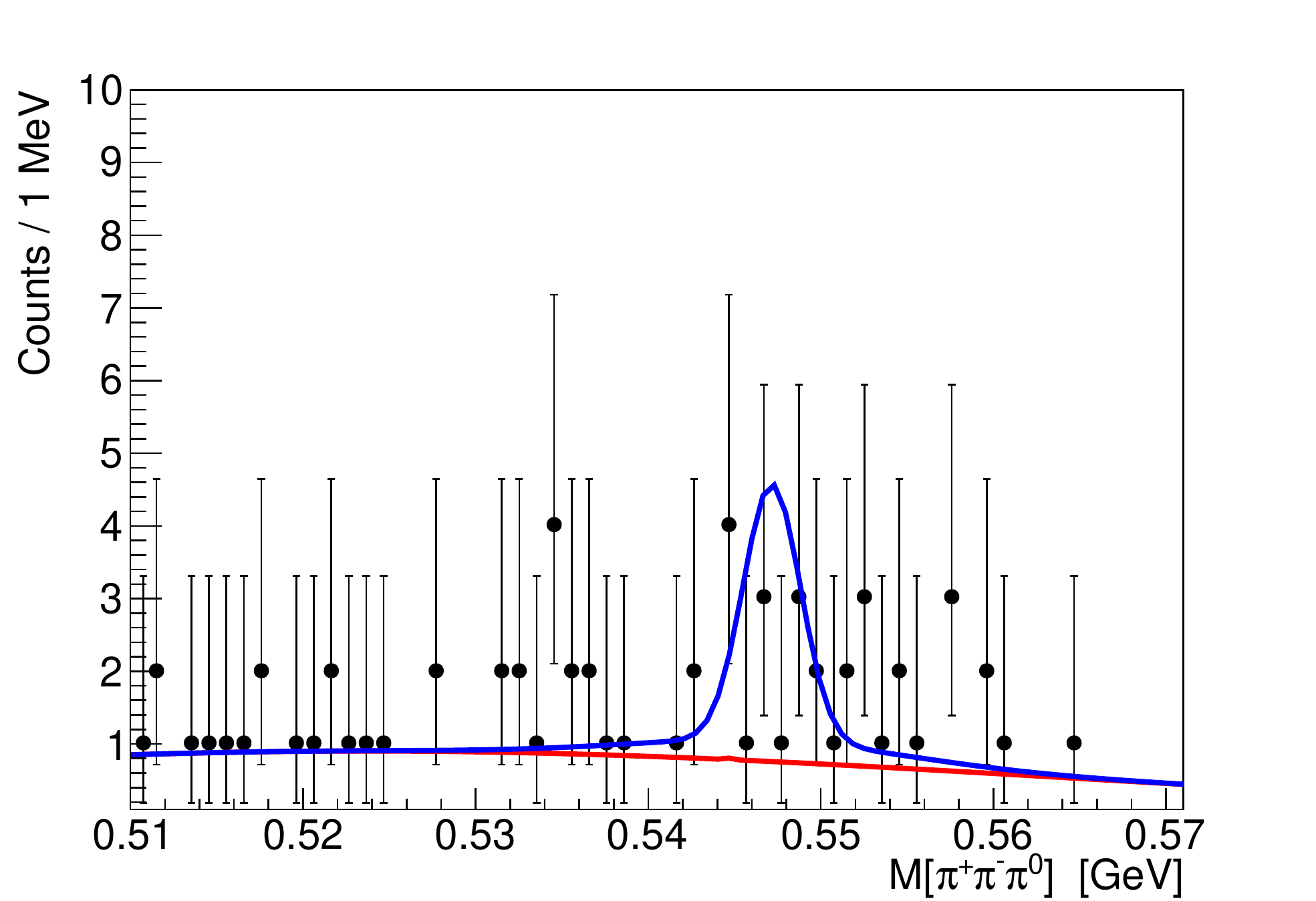}}  \\
    \end{tabular}
    \caption{ Mass distribution of  the $\eta$ candidates in $ \gamma\gamma\mu^+\mu^-$ (top), $ \gamma\gamma e^+e^-$ (middle), 
    $\pi^+\pi^-\pi^0 \mu^+\mu^-$, and $\pi^+\pi^-\pi^0 e^+e^-$ (bottom) final states. The fit function in blue represents the simultaneous fit of the four channels; the red curve shows the best-fit background component
 }
    \label{fig:Meta2}
  \end{center}
\end{figure}

\begin{figure}[htbp]
  \begin{center}
    \resizebox{1.1\columnwidth}{!}{\includegraphics{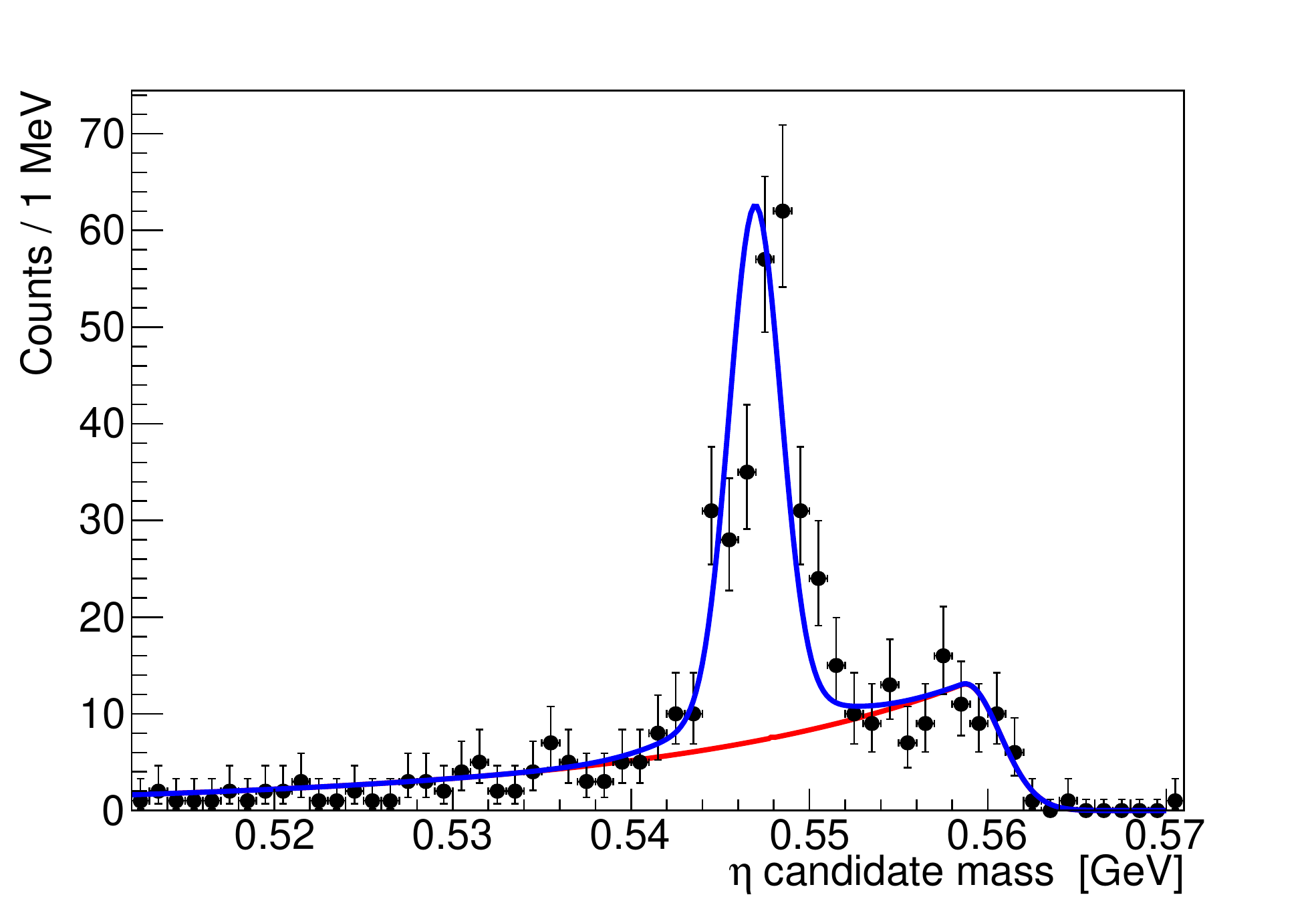}}
    \caption{$\gamma\gamma$/$\pi^+\pi^-\pi^0$ invariant mass for  $\Upsilon(2S)\to\Upsilon(1S)\eta$ 
    candidates, summing all the four final states.}
    \label{fig:mass_fitted}
  \end{center}
\end{figure}

The signal yield is extracted with a simultaneous, unbinned likelihood fit
of the $\eta$ mass distribution in four different final states with a common branching fraction, as shown in Fig. \ref{fig:Meta2}. 
For the $\eta$  invariant mass peak, we use a double Gaussian with parameters, that differ from channel to channel and fixed at  values determined by the simulation.  The background probability density function (PDF) shape, a Crystal Ball function \cite{Gaiser:1985ix} in $\eta \to \gamma\gamma$ and a Gaussian shape in $\eta \to \pi^+\pi^-\pi^0$, is chosen using the MC simulation; the parameters of the chosen PDFs, including the background yields in each channel, are left free in the fit of the four final states.
The sum of the invariant mass distributions for the four independent final states is shown in Fig. \ref{fig:mass_fitted}.


\begin{table}[h]
\begin{center}
\small
    \caption{Signal efficiencies}
    \label{tab:eff_signal}
\begin{tabular}{l c c } 
\hline \hline
                            & $ \Upsilon(1S) \to e^+e^-$ & $ \Upsilon(1S) \to \mu^+\mu^-$\\
\hline $\eta \to \gamma\gamma$     & $8.4\%$                   & $25.7\%$                         \\
            $\eta \to \pi^+\pi^-\pi^0$  & $6.4\%$                   & $7.6\%$                       \\
            $\pi^0 \to \gamma\gamma$    & $6.0\%$                   & $7.8\%$                         \\
\hline\hline
\end{tabular}
\end{center}
\end{table}

The ratio of the branching fraction for
the $\eta$ transition to that for the dipion transition is given by:
 $$
{\mathcal R}_{\eta,\pi^+\pi^-} = \frac{{\mathcal B}(\Upsilon(1S)\eta)}{{\mathcal B}(\Upsilon(1S)\pi^+\pi^-)} = 
\frac{N_{\eta, f} }{N^{\ell\ell}_{\pi\pi}}\times 
 \frac
{ \epsilon^{\ell\ell}_{\pi\pi}}
 {{\mathcal B}(\eta \rightarrow f) \cdot \epsilon^{\ell\ell}_{\eta,f}}{\rm ,} $$

 \noindent where the $\eta$ final state $f$ is  $\gamma\gamma$ or $\pi^+\pi^-\pi^0$ , $N_{\eta}$  is the signal yield from the fit and $N^{ll}_{\pi\pi}$ is the number of detected $\Upsilon(2S) \to \Upsilon(1S) \pi^+ \pi^-$ transitions for each $\Upsilon(1S) \to l^{+}l^{-}$ decay mode:  $N^{ee}_{\pi\pi} = 228167$, $N^{\mu\mu}_{\pi\pi} = 276261 $ and $N^{\rm tot}_{\pi\pi} = 504428 $. 
The charged dipion transition is selected with the same cuts as in  the $\eta \to \pi^+\pi^-\pi^0$ selection and requiring the event to have exactly one dipion  and no $\pi^0$ or $\eta$ candidates.
 According to the Montecarlo simulation the background contribution is negligible, and the efficiencies for the normalization channel are $\epsilon^{ee}_{\pi\pi}=$31.26\% and 
$\epsilon^{\mu\mu}_{\pi\pi}=$37.94\%. 
The number of $\Upsilon(2S) \to \Upsilon(1S) \eta$ events extracted from the simultaneous fit is $N_{\eta} = 241 \pm 17$.
The resulting ratio  is 
$
 {\mathcal R}_{\eta,\pi^+\pi^-} = (1.99 \pm 0.14)\times 10^{-3}
$, where the error is statistical. 

In order to study systematic effects, the ratios ${\mathcal R}_{\eta,\pi^+\pi^-}$  measured separately in four different final states are reported in Table \ref{tab:subsamples}.
The fitting procedure used for the individual channels is the
same as that used for the fit to the full sample.

\begin{table}[h]
\small
    \caption{Ratio ${\mathcal R}_{\eta,\pi^+\pi^-}$ extracted from different subsamples.}
    \label{tab:subsamples}
\begin{center}
\begin{tabular}{ l c } 
\hline\hline
  Final state & $ {\mathcal R}_{\eta,\pi^+\pi^-} {\rm ,} 10^{-3}$  \\
 \hline      $\eta \rightarrow 2\gamma \ \ \Upsilon(1S)\rightarrow \mu^+\mu^-$&$2.16 \pm 0.19$ (stat.) \\
    $\eta \rightarrow 2\gamma \ \ \Upsilon(1S)\rightarrow e^+e^-$&$ 2.15 \pm 0.38 $ (stat.)\\
 $\eta \rightarrow 3\pi \ \ \Upsilon(1S)\rightarrow \mu^+\mu^-$&$1.66 \pm 0.39 $ (stat.) \\
 $\eta \rightarrow 3\pi \ \ \Upsilon(1S)\rightarrow e^+e^-$&$1.31 \pm 0.56 $ (stat.)\\

\hline           Simultaneous fit  & $1.99 \pm 0.14 $ (stat.) \\
\hline\hline
\end{tabular}
\end{center}
\end{table}


\begin{figure}[htbp]
  \begin{center}
    \resizebox{1.1\columnwidth}{!}{\includegraphics{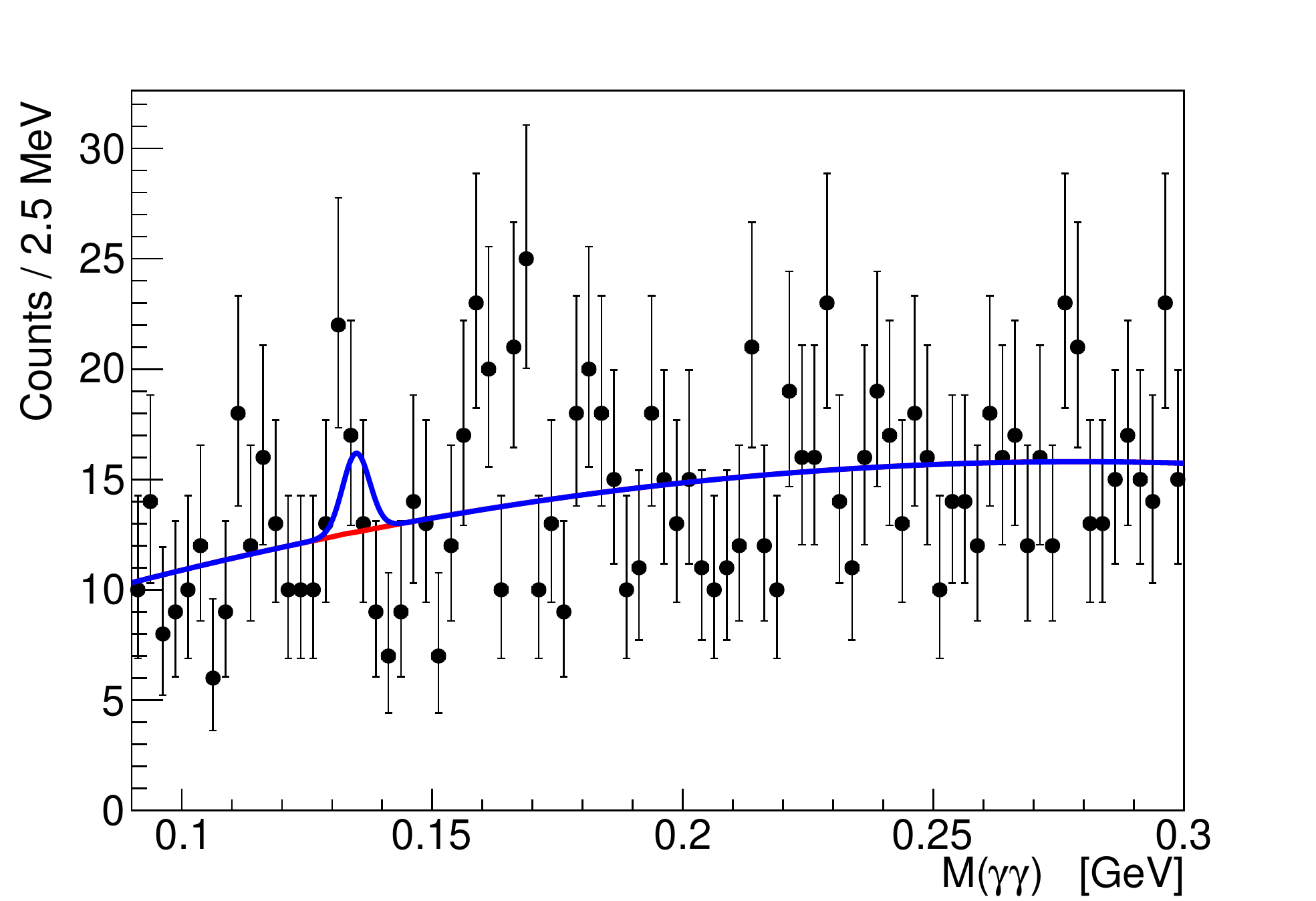}}
    \caption{Final fit to the $\gamma\gamma$ invariant mass for $\Upsilon(2S)\to\Upsilon(1S)\pi^0$ candidates.}
    \label{fig:pi0_data_tot}
  \end{center}
\end{figure}

%

\begin{table}[ht]
\begin{center}
\small
    \caption{Sources of systematic uncertainties.}
    \label{tab:systsrc}
\begin{tabular}{lcc} 
\hline \hline 

Source & $\eta$ channel &  $\pi^0$ channel\\
\hline Bhabha veto & $\pm 2.5\%$ & --- \\
       Kinematic Fit  & $\pm1.5\%$ &  $\pm1.5\%$ \\
       Background fit  & $\pm2.1\%$ & $\pm6.3\%$\\
       $\eta$/$\pi^0$ reconstruction  & $\pm4\%$ & $\pm4\%$ \\
       Signal PDF & $ \pm 2\%$  & $\pm 1.8\%$  \\
    
\hline Total & $\pm 5.7\% $ & $\pm 7.8\% $  \\
\hline
\hline
\end{tabular}
\end{center}
\end{table}


In the $\Upsilon(2S) \to \Upsilon(1S)\pi^0$ analysis, the $\gamma\gamma$ invariant mass distribution is
fitted to a Gaussian function for the signal and a third-order
polynomial for the background (Fig. \ref{fig:pi0_data_tot}).   All parameters are freely varied except the width and the mean of the Gaussian, which are set to the values determined by MC simulation. No clear evidence for a $\pi^{0}$ signal is found in either the $\Upsilon(1S)\to e^+e^-$ or the $\Upsilon(1S)\to \mu^+\mu^-$ mode. The signal yield from the fit is $N_{\pi^0} = 10 \pm 5$.

An upper limit on the number of $\Upsilon(1S)\pi^0$ candidates,  $N^{\rm UL}_{\Upsilon\pi^0}$,  is determined by generating 5000 pseudo-experiments for different values of the signal yield ranging from 0 to 30, 
using a Bayesian-frequentist hybrid approach \cite{Read:2002hq} based on the ratio of CLs between the $p$-value of the signal+background hypothesis and the $p$-value of the background-only hypothesis.
The resulting upper limit is $N^{\rm UL}_{\Upsilon\pi^0}=20.5$.

The upper limit on the  ratio $\frac{{\mathcal B}(\Upsilon(2S)\rightarrow\Upsilon(1S)\pi^0)}{{\mathcal B}(\Upsilon(2S)\rightarrow\Upsilon(1S)\pi^+\pi^-)}$ is then calculated  from the relation:
 $$
{\mathcal R}_{\pi^0,\pi^+\pi^-} = 
\frac{{\mathcal B}(\Upsilon(1S)\pi^0)}{{\mathcal B}(\Upsilon(1S)\pi^+\pi^-)} = 
\frac{N^{\rm UL}_{\Upsilon\pi^0}}{N^{\ell\ell}_{\pi\pi}}\times 
 \frac
{\epsilon^{\ell\ell}_{\pi\pi}}
 { {\mathcal B}(\pi^0 \rightarrow \gamma\gamma) \cdot \epsilon^{ll}_{\pi^0}}{\rm ,} $$
where $\epsilon^{\ell\ell}_{\pi^0}$ is the efficiency reported in Table \ref{tab:eff_signal}, reduced by a factor 1-$\sigma_{\rm sys}$=0.922 to account for systematic uncertainties.
The contributions  from each source, reported in Table \ref{tab:systsrc}, are summed in quadrature in order to obtain the final uncertainty in each channel.
The resulting upper limit is ${\mathcal R}_{\pi^0,\pi^+\pi^-} < 2.3 \times 10^{-4}$ at 90\% confidence level.


The systematic uncertainty arising from the Bhabha veto procedure is obtained comparing by the signal yields obtained with and without
the veto. The systematic uncertainty introduced by the 2S kinematic fit is studied using a 
sample of  $\Upsilon(2S) \rightarrow \chi_{b1,2} \gamma \rightarrow \Upsilon(1S)\gamma \gamma $ events that are identified by modifying the $\eta \rightarrow \gamma \gamma$ selection. 
 To reconstruct these events,  one photon in the range $90$ MeV $< E^{*} <180$ MeV and one with $380$ MeV $< E^{*} <700 $ MeV are required. 
Three different threshold values  for CL$_{2S}$  have been used to estimate the systematic 
uncertainties on ${\mathcal B}(\Upsilon(2S) \rightarrow \chi_{b1,2}\gamma  \rightarrow  \Upsilon(1S)\gamma \gamma)$. 
The systematic uncertainty due to the choice of the parameters describing the background PDF is estimated by varying each within $\pm 1 \sigma$ from the MC value in the $\eta$ transition and changing the order of the polynomial fit in the $\pi^0$ channel.
The uncertainties arising from the possible difference of signal PDF parameters between data and MC simulation are estimated by varying them within the errors and then comparing the obtained branching fractions. The uncertainties related to the track reconstruction and the total luminosity are canceled by the normalization to the $\Upsilon(2S) \to \Upsilon(1S) \pi^+\pi^-$ transition.
An
additional relative uncertainty of $4\%$ due to neutral meson
reconstruction is included.  This error is determined from the discrepancy between data and MC in $D^0 \to K^- \pi^+ \pi^0$ decay.

In summary, using  24.7 fb$^{-1}$  of data taken at  the $\Upsilon(2S) $ resonance peak energy, a measurement of the ratio ${\mathcal R}_{\eta,\pi^+\pi^-}$ is obtained:
$$
{\mathcal R}_{\eta,\pi^+\pi^-} = 
(1.99 \pm 0.14 ({\rm stat})
  \pm 0.11 ({\rm syst}))\times 10^{-3} .
$$
This result is about 17$\%$ greater than the prediction of Ref . \cite{Kuang:2006me}, 14$\%$ below the value extracted  from Ref. \cite{Voloshin:2007dx} and less than half the value predicted by scaling from the $\psi(2S) \to  J/\psi \eta $ branching fraction \cite{He:2008xk}. 
 Assuming the branching fraction ${\mathcal B}(\Upsilon(2S)\to\Upsilon(1S)\pi^+\pi^-) = (17.92 \pm 0.26)\%$ \cite{Nakamura:2010zzi}, a new measurement of ${\mathcal B}(\Upsilon(2S) \to  \Upsilon(1S)\eta)$ is obtained: 
 $$
{\mathcal B}(\Upsilon(2S)\rightarrow\Upsilon(1S)\eta) 
= (3.57 \pm 0.25 ({\rm stat})\  \pm 0.21 ({\rm syst}))\times 10^{-4}, 
$$
where an additional systematic error of  $1.4 \%$ is introduced in order to account for the uncertainties on  ${\mathcal B}(\Upsilon(2S)\to \Upsilon(1S)\pi^+\pi^-)$. This result is higher by about two standard deviations and more precise than those obtained by BaBar \cite{Babar:2011wj}, and CLEO \cite{He:2008xk}.
In addition, an upper limit for the ${\mathcal R}_{\pi^0,\pi^+\pi^-}$  ratio, a factor of four more stringent than that of CLEO \cite{He:2008xk}, is obtained:
$$
{\mathcal R}_{\pi^0,\pi^+\pi^-} < 2.3 \times 10^{-4},
$$
corresponding to the upper limit of 
$$
{\mathcal B}(\Upsilon(2S)\rightarrow\Upsilon(1S)\pi^0) < 4.1\times 10^{-5}\  ({\rm 90\%\ C.L.}).
$$
The upper limit for the ratio
$$ \frac{{\mathcal B}(\Upsilon(2S)\rightarrow\Upsilon(1S)\pi^0)}
{{\mathcal B}(\Upsilon(2S)\rightarrow\Upsilon(1S)\eta)} < 0.13\ ({\rm 90\%\ C.L.}) $$ 
is  slightly  below  the expected value of $0.16 \pm 0.02$ \cite{He:2008xk}.

We thank the KEKB group for excellent operation of the
accelerator, the KEK cryogenics group for efficient solenoid
operations, and the KEK computer group and
the NII for valuable computing and SINET4 network support.  
We acknowledge support from MEXT, JSPS and Nagoya's TLPRC (Japan);
ARC and DIISR (Australia); NSFC (China); MSMT (Czechia);
DST (India); INFN (Italy); MEST, NRF, NSDC of KISTI, and WCU (Korea); MNiSW (Poland); 
MES and RFAAE (Russia); ARRS (Slovenia); SNSF (Switzerland); 
NSC and MOE (Taiwan); and DOE and NSF (USA).


\begin{thebibliography}{99}


\bibitem{Brambilla:2010cs} 
  N.~Brambilla {\it  et al.},
  Eur.\ Phys.\ J.\ C {\bf 71}, 1534 (2011).
    
\bibitem{Gottfried:1977gp} 
  K.~Gottfried,
  Phys.\ Rev.\ Lett.\  {\bf 40}, 598 (1978).

\bibitem{Yan:1980uh}
  T.~-M.~Yan,
  Phys.\ Rev.\  D {\bf 22}, 1652 (1980).

\bibitem{Voloshin:2007dx}
  M.~B.~Voloshin,
  Prog.\ Part.\ Nucl.\ Phys.\  {\bf 61}, 455 (2008).
 
\bibitem{Kuang:2006me}
  Y.~-P.~Kuang,
  Front.\ Phys.\ China {\bf 1}, 19 (2006).

\bibitem{He:2008xk}
  Q.~He {\it et al.} (CLEO Collaboration),
  Phys.\ Rev.\ Lett.{\bf 101}, 192001 (2008).

\bibitem{Nakamura:2010zzi}
  J.~Beringer {\it et al.} (Particle Data Group), 
  Phys.\ Rev.\ D {\bf 86}, 010001 (2012).


\bibitem{Babar:2008bv}
  B.~Aubert {\it et al.} (BABAR Collaboration),
  Phys.\ Rev.\  D {\bf 78}, 112002 (2008).

\bibitem{Babar:2011wj} 
  J.~P.~Lees {\it et al.}  (BABAR Collaboration),
  Phys.\ Rev.\ D {\bf 84}, 092003 (2011).

 \bibitem{Simonov:2008sw} 
  Y.~A.~Simonov and A.~I.~Veselov,
  Phys.\ Lett.\ B {\bf 673}, 211 (2009).
  
  



\bibitem{kekb}
S.~Kurokawa, E.~Kikutani,
Nucl. Instrum. Meth.\ A {\bf 499}, 1 (2003).
and other papers included in this volume.
  
 
\bibitem{Abashian:2000cg}
  A.~Abashian {\it et al.},
  Nucl.\ Instrum.\ Meth.\ A {\bf 479}, 117 (2002).
 
  
\bibitem{Lange:2001uf}
  D.~J.~Lange,
  Nucl.\ Instrum.\ Meth.\ A {\bf 462}, 152 (2001).

  
\bibitem{Jadach:1999vf}
  S.~Jadach, B.~F.~L.~Ward, Z.~Was,
  Comput.\ Phys.\ Commun.\  {\bf 130}, 260 (2000).

 
  
\bibitem{Geant3}
  R. Brun {\it et al.}, GEANT3.21, CERN Report DD/EE/84-1 (1984).
  

\bibitem{Barberio:1993qi}
  E.~Barberio, Z.~Was,
  Comput.\ Phys.\ Commun.\  {\bf 79}, 291 (1994).
  
  
\bibitem{Gaiser:1985ix} 
  J.~Gaiser {\it et al.},
  Phys.\ Rev.\ D {\bf 34}, 711 (1986).

\bibitem{Read:2002hq} 
  A.~L.~Read,
  J.\ Phys.\ G {\bf 28}, 2693 (2002).
  
 

   
\end{thebibliography}
\end{document}